# Propagation of disruptions in supply networks of essential goods: A population-centered perspective of systemic risk

William Schueller[1,2,*], Christian Diem[1,3,*], Melanie Hinterplattner[4], Johannes Stangl[1], Beate Conrady[5,1], Markus Gerschberger[6,4] and Stefan Thurner[2,1,7]

[1] Complexity Science Hub Vienna, Josefstädter Strasse 39, 1080 Vienna, Austria
[2] Medical University of Vienna, Spitalgasse 23, 1090 Vienna, Austria
[3] Vienna University of Economics and Business, Welthandelsplatz 1, 1020 Vienna, Austria
[4] University of Applied Sciences Upper Austria, Logistikum, Wehrgrabengasse 1-3, 4400 Steyr, Austria
[5] University of Copenhagen, Department of Veterinary and Animal Science, Grønnegaardsvej 8, 1870 Denmark
[6] Josef Ressel Centre for Real-Time Value Network Visibility, Logistikum, University of Applied Sciences Upper Austria, Wehrgrabengasse 1-3, 4400 Steyr, Austria
[7] Santa Fe Institute, 1399 Hyde Park Road, Santa Fe, NM 85701, USA

## ABSTRACT

The Covid-19 pandemic drastically emphasized the fragility of national and international supply networks (SNs), leading to significant supply shortages of essential goods for people, such as food and medical equipment. Severe disruptions that propagate along complex SNs can expose the population of entire regions or even countries to these risks. A lack of both, data and quantitative methodology, has hitherto hindered us to empirically quantify the vulnerability of the population to disruptions. Here we develop a data-driven simulation methodology to locally quantify actual supply losses for the population that result from the cascading of supply disruptions. We demonstrate the method on a large food SN of a European country including 22,938 business premises, 44,355 supply links and 116 local administrative districts. We rank the business premises with respect to their criticality for the districts' population with the proposed systemic risk index, $SRI^{crit}$, to identify around 30 premises that—in case of their failure—are expected to cause critical supply shortages in sizable fractions of the population. The new methodology is immediately policy relevant as a fact-driven and generalizable crisis management tool. This work represents a starting point for quantitatively studying SN disruptions focused on the well-being of the population.

*The authors acknowledge the equal contributions of WS and CD.

**INTRODUCTION**

In modern economies the production of most goods and services depends on large, complex supply networks (SN) consisting of numerous tightly interwoven supply chains (Choi *et al.*, 2001; Gerschberger *et al.*, 2012; Wiedmer & Griffis, 2021). These networks can be highly fragile, meaning that the disruption of crucial supply links can lead to critical production failures, even in distant regions and industries, causing high business related and overall economic costs (Pinior *et al.*, 2012a; Ivanov & Dolgui, 2020). There are several examples where a disruption at a single company (or even production site) lead to global problems in production, probably the most prominent example being the lack of semiconductors for car manufacturing (Burkacky *et al.,* 2021).

Recently, the Covid-19 pandemic illustrated the consequences of supply disruptions at a new level of severity, threatening the survival of companies across all sectors and industries at a global scale (Craighead *et al.*, 2020; Wenzel *et al.*, 2020). Rapid supply shocks and demand swings presented considerable managerial challenges and forced managers to implement response plans under persistent uncertainty to protect their firms and maintain supply levels (Bode *et al.*, 2021). The Covid-19 pandemic also demonstrated that pandemics are qualitatively different from typical supply disruptions and thus require a different type of response (Craighead *et al.*, 2020). Consequently, research started to examine various pandemic response capabilities and strategies (Ketchen & Craighead, 2020; van Hoek, 2020; Bode *et al.*, 2021) that assist managers to better prepare for future pandemics and crises.

Yet, not only companies are affected by disruptions in the supply network, also the supply of people with essential goods—such as food, hygiene, and medicine—is heavily reliant on the functioning of SNs. While largely overlooked in the supply chain management (SCM) literature (except for humanitarian and disaster relief/recovery studies), the functioning of SNs of essential goods is of tremendous societal importance. In fact, disruptions in the production and distribution of essential goods can lead to significant shortages in basic supply, thus affecting the well-being of millions (Mollenkopf *et al.*, 2020).

Hindsight shows that while managers were struggling to keep operations and supply chains up and running during the pandemic, governments were under pressure identifying the critical players that ensure the supply of people with essential goods. In order to develop future contingency plans, policy makers have to address questions like: "Which actors carry high systemic risk for the supply of the population?", "If a disruption materializes somewhere, how strong will it affect other SC members and eventually the population?", and "which part of the population will be (most) affected?" (Pinior *et al.,* 2012b; Pinior *et al.,* 2015; European Commission, 2021). However, up to now, a lack of data and a shortfall of existing quantitative methodology makes answers difficult. Studying the effects of supply disruptions of essential goods on the population and developing an assessment tool for systemic risks in SNs to guarantee individual and community well-being in times of crisis is thus both timely and relevant. Notably, the areas of application for a systemic risk assessment tool extend well beyond the Covid-19 pandemic. The increasing number of disruptions in today's globally interconnected economy makes it imperative to better understand, manage, and protect critical SNs.

We address this gap by studying the propagation of disruptions and developing a practical systemic risk assessment tool for governments, civil defense/protection, and decision makers to spot and quantify systemic risks in SNs. Specifically, we develop an economically and mathematically founded simulation methodology for systematically quantifying the effects supply network failures have on the supply of essential goods for the population. We empirically demonstrate our risk assessment tool based on a large and unique real-world food supply network. Our tool enables decision makers to identify the systemically critical players, quantify the impact of their hypothetical default, and to assess how a severe initial disruption (like production failures, factory or border closures, natural disasters) would propagate and percolate through the network and how it finally will affect the population. Thereby, we expand and enrich theory and methodology of disruption propagation. The newly assembled data set goes beyond the commonly studied dyadic buyer-supplier relationships to include all relevant players of a specific food supply network–from primary production to the final consumers, i.e., the population. As such, it allows one of the most comprehensive empirical analysis of real-world SNs to date. A unique aspect of this work is the emphasis on the importance of incorporating the entire population of a country—at the level of fine grained local administrative districts—as an integral part of the supply chain and its risk management. We will also discuss and outline how supply network research can drive transformation within society.

## LITERATURE REVIEW

We identified three literature streams as especially relevant for studying disruption propagation effects and developing the decision-making tool: supply networks, supply disruptions and their propagation through networks, and systemic risk (indices).

### Supply networks

Supply networks are an integral part of supply chain management. In fact, "firms practicing supply chain management have the potential to leverage their networks for the exchange of resources, goods, knowledge, and services toward the objective of serving the final customer" (Autry & Griffis, 2008, p.159). Supply chain management literature shows that SNs are critical to achieve competitive advantage (Basole & Bellamy, 2014), because they affect the production of goods and services (Borgatti & Foster, 2003), financial and operational performance (Kim *et al.,* 2011; Lu & Shang, 2017), innovation (Sharma *et al.,* 2020; Potter & Wilhelm, 2020), as well as resilience to disruptions (Nair & Vidal, 2011; Kim *et al.,* 2015; Wiedmer *et al.,* 2021). A network is generally defined as "a set of nodes and the set of ties representing some relationships, or lack of relationship, between the nodes" (Brass *et al.,* 2004, p. 795). Accordingly, a SN refers to the "*'exchange relationships' between suppliers, customers, and their partner firms that are necessary for manufacturing and providing goods and services to the market.*" (Wiedmer & Griffis, 2021, p. 265) Note, the terms supply network and supply chain network are commonly used interchangeably (e.g., Falcone *et al.,* 2021; Li *et al.,* 2021).

Although there exists wide agreement that firm performance depends not only on a dyadic buyer-supplier relationship, but rather on the performance of the entire SN, entire networks are often not of explicit concern—neither for practitioners nor for academic scholars (Gualandris *et al.,* 2021). This is, in part, due to the fact that complete SN analysis requires data that is often either impossible to obtain, or—if technically possible—prohibitively expensive (Knoke & Yang, 2019). Consequently, in research SNs were assumed as either simplified real-world models, random networks, simulated networks intended to represent general instances of specific network topologies of SNs, or—commonly in the field of supply chain management—as subsets of whole SN structures. For example, Choi & Hong (2002) used interview and archival data to map three SNs that pertain to one component in the automotive industry. Using the same data set, Kim *et al.* (2011) demonstrate how social network analysis can be used to investigate the structural characteristics of SNs. Shao *et al.* (2018) analyzed the SN of Honda Motor up to the fourth level and used social network centrality measures to identify nexus suppliers. Based on a data set of 219 first-tier suppliers within the Toyota SN, Potter & Wilhelm (2020) investigated how the structure of the SN influences each firm's ability to form supplier–supplier innovations with other network partners. Other recent network studies examine first-tier SNs to investigate the impact of supply base structural complexity (Lu & Shang, 2017) and supply base R&D intensity (Dong *et al.,* 2020) on firm financial performance.

Recently, Choi *et al.* (2021) re-emphasized that supply chains are complex adaptive networks (Choi *et al.,* 2001) that need to be analyzed beyond the first- (and second-) tier level. Yet, to date there exists limited empirical work that examines real-world SNs, with some notable exceptions. For instance, Wiedmer & Griffis (2021), who study the network topologies of 21 extended five-tier SNs and Gualandris *et al.* (2021) who study 187 extended supply chains. Other studies in related fields investigate nation-wide supply networks that are reconstructed from transaction level value added tax records (Borsos & Stancsics, 2020;

Dhyne *et al.*, 2021; Diem *et al.,* 2021), survey data (Inoue & Todo, 2019), or business mobile phone communication data (Reisch *et al.*, 2021).

**Supply disruptions and network propagation**

Given their frequency and severity, supply (chain) disruptions are a significant managerial concern (Bode & Macdonald, 2017; Polyviou *et al.,* 2020). Supply disruptions have been defined as "unplanned and unanticipated events that disrupt the normal flow of goods and materials within a supply chain […] and, as a consequence, expose firms within the supply chain to operational and financial risks" (Craighead *et al.,* 2007, p.132). Supply disruptions may decrease performance related to operations (e.g., stockouts, production shutdowns), finance (lost sales, premium freight charges), and relationships (Hendricks & Singhal, 2005; Wu *et al.,* 2007; Hendricks *et al.,* 2009; Bode & Macdonald, 2017). To avoid such negative consequences, SCM literature proposes several capabilities, such as redundancy, flexibility, visibility, and collaboration (Christopher & Peck, 2004; Sheffi & Rice, 2005; Pettit *et al.,* 2013), as well as pro- and reactive response strategies, such as buffering and bridging (Bode *et al.,* 2011; Vanpoucke & Ellis, 2020) that firms can use to withstand or quickly recover from disruptions.

Another stream of research investigates how supply disruptions propagate through the network and the role the network structure has. For example, Bode & Wagner (2015) investigate how the structure of firms' upstream supply chains affect the frequency of disruptions. Based on the results of a survey including 396 firms they find that horizontal, vertical, as well as spatial complexity increase the frequency of disruptions. In another study, Li *et al.* (2021) use a simulation approach to examine downstream and upstream propagation of disruptions in the SN. They apply their study to a data set with 121 nodes and 193 links (unweighted), containing the tier 1 and tier 2 suppliers of Toyota and Honda derived from Bloomberg SPLC database. Results highlight the importance of differentiating between forward and backward disruption propagation as they have different effects on network and individual firm performance. Li & Zobel (2020) investigated the resilience of SNs in the presence of disruptions. They use an epidemiological SIR (susceptible infected recovered) model to simulate the disruption propagation for simulated networks (scale-free, small world, Erdös-Reny random network) of sizes with 100, 300 and 500 nodes. Findings indicate, among others, that resilience is affected more by node risk capacity than by the topology of the three types of simulated networks.

Conceptually related to our present study, Burgos & Ivanov (2021) investigate the effects of the Covid-19 pandemic on the food supply chain by means of a so-called digital twin (a very detailed agent-based model). They consider 28 supermarket locations, 3 distribution centers and 10 different product categories that are supplied by a sample of 30 suppliers (3 per product). Their simulation results show that surges in demand and supplier shutdowns have had the highest impact on operations and performance, whereas the impact of transportation disruptions was rather low. A recent literature review concerning disruption propagation in SNs is provided by Dolgui *et al.* (2018). They identify different methodologies to study disruption propagation including mathematical optimization, simulation (system dynamics, agent-based modelling, event simulation), control-, and reliability-theory.

Due to data availability constraints, analyzed SNs mostly contain only a few hundred firms and have been commonly represented as unweighted networks. This means that a supply relation between two firms either exists or not; Unweighted links ignore the actual importance of different relationships and treat them all as equally important. Recent advances in data availability and quality—large scale networks with monetary volumes of supply links and industry information—have led to the development of network-based system risk indices that make use of this crucial additional information (Diem *et al.,* 2021).

**Systemic risk (indices)**

The concept of disruption propagation is closely related to the stream of literature on systemic risk in complex networks. The term *systemic risk* refers to the risk that the failure of a single node—or a small set of nodes—exposes a large part of the network (system) failing to perform its regular function. 'Failing to perform its regular function' can have different meanings in different types of networks (systems). For example, the functioning of the communication network was studied by removing network nodes—in a random or targeted way—and the effect on the network's functioning was measured by characteristics such as the network diameter and the size of the largest connected component (Albert *et al.,* 2000). Similarly, the functioning of power grids in response to the failure of single nodes has been examined by simulating cascading failures in the network (Crucitti *et al.,* 2004). The systemic risk in financial networks has been investigated by simulating the propagation of bank defaults via interbank loan contracts between banks (Eisenberg & Noe, 2001; Boss, Summer & Thurner, 2004).

Especially since the great turmoil of the 2008 financial crisis, being able to measure the systemic risk of individual banks became increasingly relevant to better inform decisions of regulators and governments (Glasserman & Young, 2016). One way to do so is to simulate—based on economically meaningful shock spreading mechanisms—how the financial losses—triggered by an initial failure of a bank—propagate via the financial connections between banks and then aggregate the losses suffered by all banks to a single number, i.e., the systemic risk index of the bank. Examples for such systemic risk indices are the Contagion Index (Cont *et al.,* 2010; Cont *et al.,* 2013) and DebtRank (Battiston *et al.,* 2012; Thurner & Poledna, 2013) for measuring systemic risk in direct financial exposure networks and the Indirect Contagion Index (Cont & Schaanning, 2019) for measuring systemic risk in indirect financial exposure networks.

There are few studies using standard network centrality measures to assess systemic risks in supply networks (Ledwoch *et al.*, 2016). In contrast to financial networks, however, literature on the development of corresponding systemic risk indices for firms in large scale supply and production networks is still rare. Notable examples are the application of DebtRank to a large-scale supply network data set of Japan (Fujiwara *et al.,* 2016) and the development of the economic systemic risk index (ESRI) and its application to the Hungarian supply and production network (Diem *et al.,* 2021). So far, the SCM literature deals with systemic risk mostly on a conceptual and non-quantitative level (Scheibe & Blackhurst, 2019).

To conclude, risks imposed by disruptions in the supply network have been studied extensively, yet SCM literature focuses almost exclusively on examining these risks from the perspective of a focal firm and related to losses in business performance. Recently, Mollenkopf *et al.* (2020) emphasized the importance of treating the population as an integral part of supply chain management, calling for more research on individual and community well-being (Mollenkopf *et al.,* 2021). This study enriches SCM literature by empirically examining systemic risk and disruption propagation effects in an extended food supply network in order to ensure population well-being.

**METHODOLOGY**

The main goal of this work is to develop an economically and mathematically founded simulation methodology that allows us to simulate the propagation of supply chain disruptions, starting from any given point in the network, downstream until reaching the population. Further, we develop a systemic risk index to define the criticality of each actor in the SN. The method is able to identify those actors that carry high systemic risk for the supply of the population—modelled at the level of local administrative districts—with essential goods and to estimate how strong a given disruption will affect the population locally; in particular also which parts of the population will be affected and how severely. Finally, we apply our simulation method to a large real-world SN data set (including population data as well as all relevant upstream suppliers, up to raw material suppliers) and show how the population is exposed to shortages in basic supplies stemming from disruptions in the supply network.

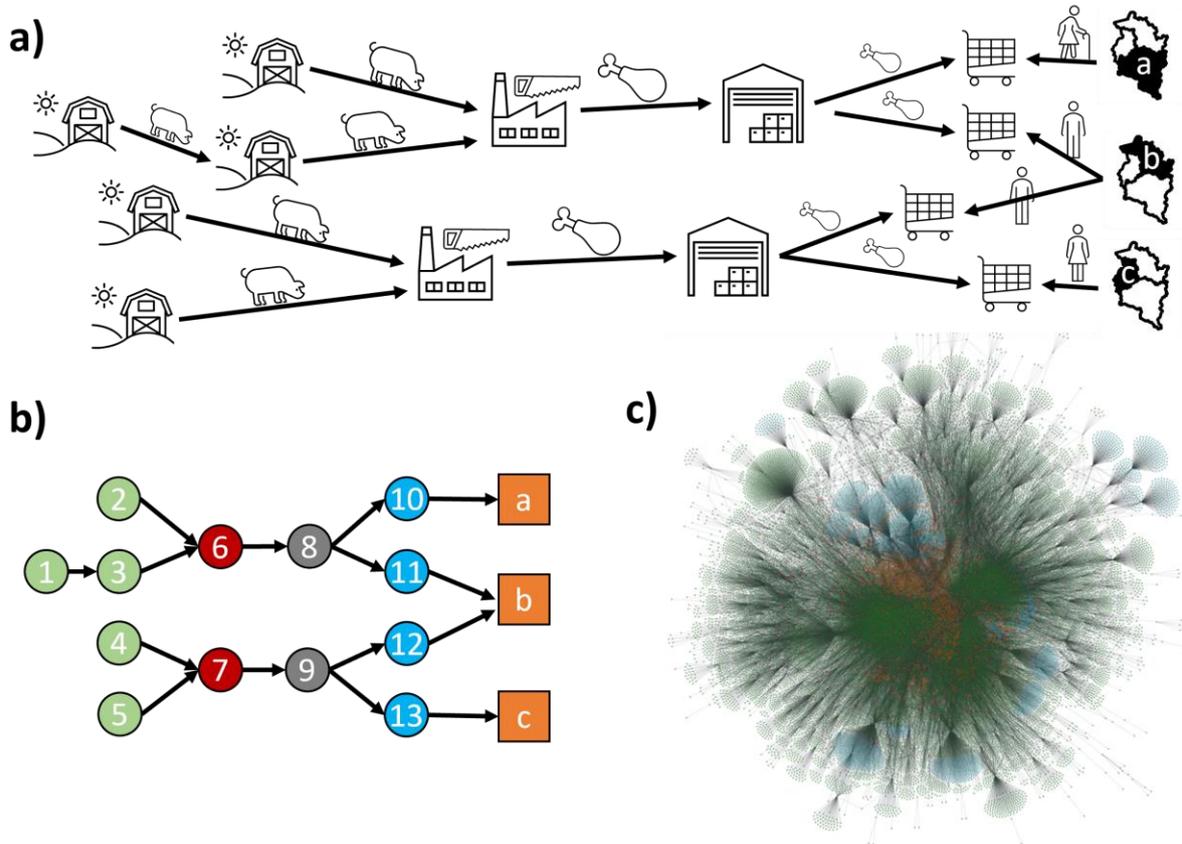

**Figure 1.** Visualization of the Austrian pork supply network.

a) Schematic representation of the considered types of actors and how they are connected. Different types of primary production, slaughterhouses and meat processors, distribution centers of supermarket chains, local supermarkets, and administrative districts (population nodes). Goods flow downstream to stores where the local population buys them.

b) Network abstraction of the schematic example in a). Node color indicates business premise types, edges indicate flows of products between the nodes; round symbols represent supply network nodes, square symbols represent population nodes. In the implementation also node sizes and edge weights are considered; they are omitted here for the clarity of presentation.

c) Visualization of the reconstructed Austrian pork supply network from the collected data, containing approximately 23,000 nodes. A core periphery structure is visible, where farms and supermarkets are located at the periphery; slaughterhouses, meat processors, and distribution centers are close to the center of the network.

**Network Representation**

We model the supply network of essential goods as a directed and weighted complex network, where the business premises of firms are the nodes in the network and the links (or edges) represent the flow of goods between those business premises (nodes). Edge weights correspond to the volume of goods delivered from the supplier to the buyer in a given time period. We consider only one type of good per node due to the nature of the collected data. With data available, the method can be extended straightforwardly to multiple types of goods per node. We embed the local population into the supply network by modelling the administrative districts in the country as population nodes. These are directly linked to the supermarkets located within the respective district, where the local population satisfies their demand for basic supplies. In the model the local population in every district is exposed to supply disruptions that occur anywhere in the SN upstream of the local supermarkets.

We illustrate the network model in Figure 1 with the Austrian pork supply network. Panel a) shows a schematic representation of the network structure, starting from primary production of pigs that are delivered to slaughterhouses, which, in turn, deliver processed meat to the distribution centers of large supermarket chains. The processed meat finally is shipped to the supermarkets (shopping carts). People living in three different districts (a, b and c) buy the meat at the supermarkets within the respective district. Figure 1 b) shows the abstraction of the supply chain as a complex network, where color corresponds to the different types of business premises in the SN (for ease of illustration we omitted link weights). Round nodes correspond to business premises, square ones represent districts (i.e., the population). Figure 1 c) shows the empirical pork SN of Austria with roughly 23,000 nodes (without population nodes). The underlying data set is described in detail in the data section.

**Simulation study design**

The design of our simulation study includes five steps:
1. observing the *undisrupted* supply network,
2. assuming (temporary) *disruption* of a specific node,
3. simulating how this *node's disruption propagates* downstream,
4. calculating for each population node the *remaining local product supply*,
5. calculating the fraction of population below a *critical supply threshold*.

First, we consider the undisrupted SN, where each node in the network can deliver its regular amount of products to its buyers, as illustrated in Figure 1 b). Second, we assume the (temporary) full operational disruption (initial failure) of one specific node in the network. In the example network we assume the disruption of a slaughterhouse (*node 7*), illustrated in Figure 2 a) with a red cross. Third, we assess the importance of the slaughterhouse for the population, by simulating its disruption impact on nodes that are located downstream in the supply network down to the population nodes. The disruption propagates to *node 9* (a distribution center) directly and *nodes 12* and *13* (supermarkets) indirectly. Consequently, the population nodes

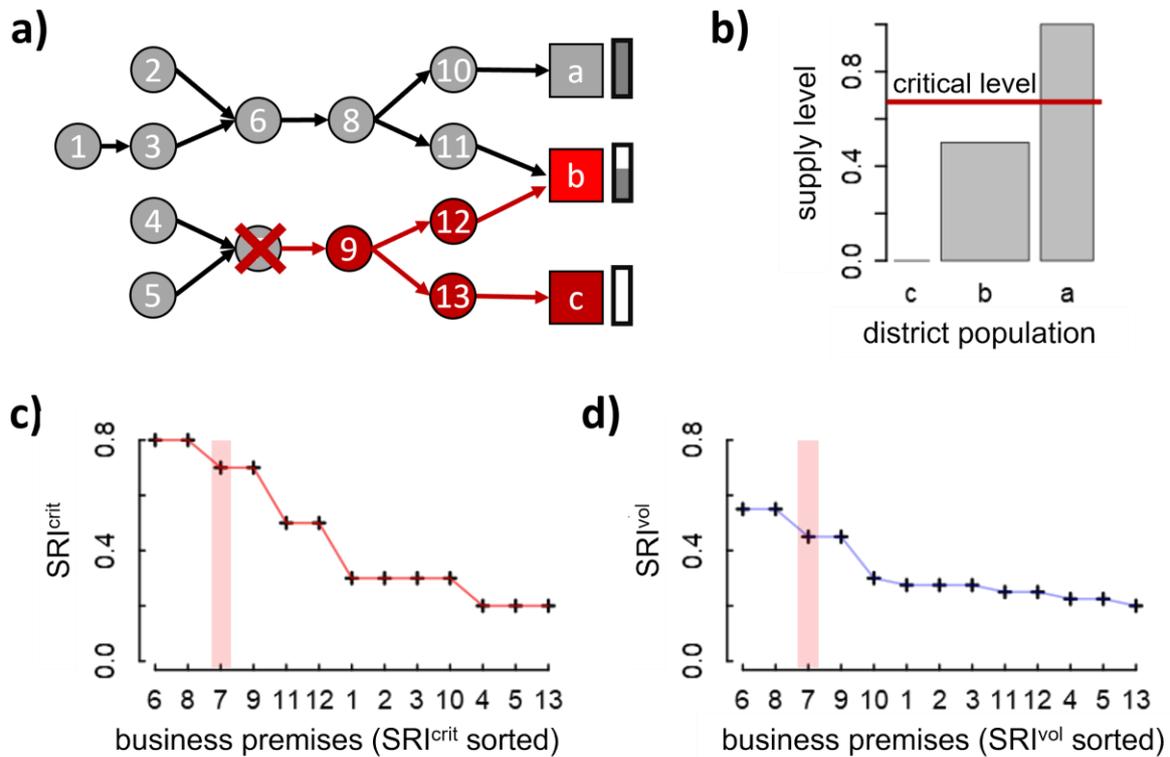

**Figure 2.** Illustration of systemic risk assessment for individual business premises.

a) Illustrates the temporary disruption (e.g., fire, forced closure by authorities) of *node 7* in the example supply network, marked with a red cross. The nodes affected downstream of *node 7* are colored in red. Therefore, the population districts *c* and *b* suffer a supply loss of 1 (100%) and 0.5 (50%), respectively, whereas district *a* is not affected by any supply loss. The remaining supply is indicated by the height of the small grey bars next to districts.

b) Bar chart visualizing the remaining supply level for each district after the failure of *node 7*. The most affected district is on the left; bar width corresponds to population size. The red horizontal line indicates the critical supply threshold of 66% (0.66). Therefore, the initial failure of *node 7* affects 70 % of the population criticality ($SRI^{crit}$ = 0.7) and reduces the total volume of the pork meat supply to the population by 45% ($SRI^{vol}$ = 0.45).

c) Shows the ranking of all 13 supply network nodes according to their systemic risk index, $SRI^{crit}$. The x-axis shows the node IDs sorted according to their, $SRI^{crit}$ values—most risky premises are on the very left. The y-axis denotes the $SRI^{crit}$ of each node (fraction of population that is critically affected). The initially failing *node 7* has rank 3 (red shading).

d) Shows the ranking of all 13 nodes according to the systemic risk index, $SRI^{vol}$. Again, node *7* has rank 3, highlighted with red shading. Note the difference in the rankings.

*b* and *c* suffer from a reduction in their basic supply because their local supermarkets have been disrupted, whereas the population in *node a* (district *a*) is not affected. Fourth, after simulating the downstream disruption propagation, the remaining supply levels for each of the three districts *a, b* and *c,* can be calculated. Figure 2 b) shows that district *a* is not affected and remains with 100% of its original supply, whereas the supplies for districts *b* and *c* dropped to 50% and 0%, respectively. We assume that the districts *a*, *b* and *c* have relative shares of the population of 30%, 50% and 20%, respectively—in Figure 2 b) these population shares are illustrated by the bar width. Fifth, we set a critical threshold for the basic supply level of the population, below which the population's well-being is seriously affected. Based on this threshold and the population shares it can be assessed how large the fraction of the population is that suffers from seriously reduced well-being in response to the initial shock. Here we choose the critical threshold to be 66% of the undisrupted full supply level—illustrated by the red horizontal line. In our example districts *b* and *c* fall below this critical threshold. Therefore, 70 % of the population suffers a critical loss in their essential supply when *node 7* fails.

Based on the fraction of the population below the critical threshold we define the *systemic risk index*, $SRI^{crit}$. In our example, the value of 0.7 (70%) corresponds to the *systemic risk index with respect to the critical supply of the population* of *node 7*, i.e., $SRI_7^{crit} = 0.7$. Accordingly, the value $SRI^{crit}$ can be also calculated for the other 12 supply nodes in the network. Figure 2 c) shows the ranking of all 13 nodes based on their systemic risk index, $SRI^{crit}$. The most risky nodes are to the very left, the least risky nodes to the right. *Node 7* is ranked as the third most risky node and is highlighted with a red shading.

Alternatively, we can assign to each node a volume-based *systemic risk index*, $SRI^{vol}$. It is calculated as the *total volume of essential supply to the population that is reduced when a node's operation is fully disrupted*—this corresponds to the sum of 1 minus the height of the bars multiplied by the bar width in Figure 2 c). In our example, *node 7* has a $SRI_7^{vol} = 0.45$ (calculated as $0.2 \cdot (1-0) + 0.5 \cdot (1-0.5) + 0.3 \cdot (1-1)$), where, districts *c*, *b* and *a* have populations of shares of 20%, 50% and 30%, respectively; the terms $(1-0)$, $(1-0.5)$ and $(1-1)$ correspond to the supply losses of districts *a*, *b* and *c*. Figure 2 d) depicts the ranking of the 13 supply network nodes with respect to the volume based systemic risk index, $SRI^{vol}$. *Node 7* is highlighted again with the red shading. Note that the two rankings differ. The mathematically exact formulations of $SRI^{crit}$ and $SRI^{vol}$ are provided in Appendix A.

**Population-centered systemic risk index**

Setting the critical supply threshold captures the population well-being in the systemic risk index, $SRI^{crit}$. In contrast, a volume-based formulation, $SRI^{vol}$, focuses mostly on economic consequences, but ignores the fraction of the population, suffering from a serious SN disruption. This can be illustrated in the following example. Consider that a large supplier, accounting for 10% of the supply of a specific product type across all districts in the country, is disrupted. Then, this disruption is large in volume and affects every district, but since the supply loss is spread evenly across the country, arguably nobody's well-being will be critically affected. A simple volume-based measure cannot distinguish if 10% in volume is reflected in 10% of the population having no supply at all, or if 10% in volume means that 100% of the population face a mere 10% loss.

This raises the question of how to define a proper critical supply threshold. Ideally, it is chosen such that it can be interpreted as the basic supply level under which a serious loss in the population's well-being occurs. It could be either formulated in terms of a minimum level of calories per day or nutrition values that are deemed medically necessary, or qualitatively, by surveying the population to find what levels are tolerable for different types of essential goods. The proposed systemic risk index, $SRI^{crit}$, allows us to rank firms and business premises in the SN with respect to their likely impact on the essential product supply and thus well-being. We discuss the advantages of the formulation of the systemic risk index, $SRI^{crit}$, and its implications for decision makers in the conclusion section.

**Simulation of downstream disruption propagation**

We shortly summarize the features of the employed downstream disruption propagation algorithm here and provide a more detailed overview in Appendix C. The algorithm captures the essential dynamics of real-world disruption propagation. For each supplier-buyer relation (supply link) the propagation mechanism considers the importance of the supplier for the buyer's supply by comparing delivery volumes across suppliers for a given product. A supplier A delivering 50% of a product to a warehouse can cause a larger disruption for the warehouse than a supplier responsible for only 10% of the product's supply. Note that if link weights are not known, the algorithm assumes equal importance. This method can be readily calibrated to available SN data. The regular undisrupted output level of nodes is inferred from the weighted out-degree of nodes, i.e., the sum of all delivered goods. The regular supply level of each population node is inferred as the sum of all weighted in-degrees of the supermarkets within one district, i.e., all products that are delivered to the supermarkets in the respective district.

# EMPIRICAL INVESTIGATION OF DISRUPTION PROPAGATION AND SYSTEMIC RISK IN NATION-WIDE PORK SUPPLY CHAINS

**Data**

For a meaningful illustration of the above methodology, we need a detailed data set that allows for a sufficiently realistic reconstruction of a country-wide supply network and that covers the majority of all relevant nodes in the investigated sector. Since the majority of the population satisfies its demand for basic supplies—essential food and hygiene products—primarily from local supermarkets, the upstream SNs of the large supermarket chains are a suitable choice. The increasing trends for online shopping and delivery of food products can be modelled in the same ways since deliveries are conducted either from local supermarkets directly or regional distribution centers. Since it is out of scope to model the full supply network of the supermarket sector of a whole country, we focus on the pork SN as a prime example of essential food supply.

We collected data on a nation-wide (Austria) pork supply chain, ranging from primary production to consumption at the local population level. The data set provides almost full coverage and contains 22,938 business premises as supply network nodes, 44,355 supply links between the business premises, as well as the population data of 116 administrative districts (population nodes). The data set includes trade volumes between the business premises, allowing for reliable estimates of disruption propagation since buyer-supplier relations are accurately represented. In comparison, commonly used data sets in existing literature that are assembled from business intelligence databases (such as Bloomberg, CompuStat, etc.) often only contain binary information about buyer-supplier links, i.e., if a link exists or not.

The data set is composed from several data sources and was created in the course of a joint project with the Austrian Ministry of Agriculture. It unifies animal trade data, supermarket data and public records; almost all relevant nodes directly associated with the pork supply network have been identified. These include a full coverage of the primary production nodes (breeding farms, fattening farms, wholesalers, and animal transport), food processing industry (slaughterhouses, meat processors) and the distribution network (distribution centers and supermarkets) of the four large supermarket chains operating in Austria (joint market share > 90%). Thus, it almost fully covers the relevant business premises of the entire sector. In places where data was missing (because not yet collected), we resorted to imputation algorithms based on monte carlo simulations that make use of the empirical supply link volume distribution of the already collected data, overall production volumes of nodes and geographical distances between premises. For details of the individual data sets, the network construction, and the imputation procedure we refer to Appendix C.

*Considering the population in supply network models*

A unique aspect of this study is the explicit consideration of the population in the SN. Even in a small country the population cannot be modeled as a homogenous group when it comes to the supply of essential goods. Depending on regional differences, the level of urbanity, geographical topology and many other factors, the supply realities can be widely different and therefore disruptions in one part of the supply network can affect some parts of the population drastically, while others will not even notice the disruption. Therefore, for a reasonable granular assessment of the vulnerability of the population one should regionally differentiate the population into population nodes, based on e.g., administrative districts. This takes care of a large part of the heterogeneities. This differentiation can be extended to other socio-economic factors like access to a car, the financial situation, etc.

### *Relevance of the Austrian pork supply network*

For several reasons the pork SN is an interesting network to apply our methodology. The supply chain management literature focused on meat SNs already in earlier studies investigating, for example, the lean supply chain concept (Taylor, 2006; Perez *et al.,* 2010), outsourcing of processing steps (Hsiao *et al.,* 2010) or supply chain robustness (Vlajic *et al.,* 2012). From an economic perspective the sector is of considerable size even for a small country. The gross value of primary production of pig farming amounts to 831 million Euros in Austria, whereas the revenues of the slaughterhouses and meat processors in Austria amounts to 4.7 billion Euros (Austrian Ministry of Agriculture, 2021). From a population perspective the pork supply network is a prime example of an essential food supply network. It is the most important type of meat and thus contributes to a large part of people's nutrition in the country of consideration. In general pork meat is in many countries a key source of protein and thus a key component for basic food supply and consequently relevant for the population´s long term well-being.

## **RESULTS**

The data-driven simulation methodology and the data set allow us to highlight different aspects of how failures in the SN affect the basic supply of a country's population. We illustrate these with the Austrian pork SN.

### *Analyzing simulated supply losses for individual business premises*

The presented methodology allows for an in-depth analysis of the consequences when an *individual* SN node (business premise) is severely disrupted. It visualizes how the effects are distributed across the population in different districts and therefore if and where critical supply shortages are likely to occur if no action is taken. We exemplarily analyze the consequences of the hypothetical disruptions of a distribution center of a large supermarket chain—illustrated in Figure 4 a) and Figure c)—and a large slaughterhouse—illustrated in Figure 3 b) and Figure 3 d). Due to anonymization and data protection agreements, specific business premises cannot be named. To visualize the supply losses to the local population in response to the disruption we show a map where every district is colored according to its supply losses. Districts with low supply levels are colored in yellow, districts with high levels in dark blue.

In case of the failure of the selected distribution center, the map in Figure 3 a) clearly shows a very localized, but sizable (yellow) effect on the population of in the capital of Austria, Vienna. This is due to the fact that geographically narrow regions—often determined by the boundaries of federal states—are supplied from only one specific distribution center for a given supermarket chain. The supermarket sector in Austria is highly concentrated and the market shares and relative share of supermarkets vary among federal states. Therefore, the disruption of a distribution center with a high market share and that supplies a highly populated state can cause a critical supply loss to a large part of the population. Even within the affected state individual districts are affected to different degrees. This is explained by varying market shares across districts for the supermarket chain operating the chosen distribution center.

The disruption propagation triggered by the selected slaughterhouse—visualized in Figure 3 b)—clearly shows that no district suffers severe supply shortages, but different districts across the country are likely to be affected by the disruption (lighter blue tones). The reason for this is that large slaughterhouses typically sell to many different meat processors that, in turn, often supply different supermarket chains and usually warehouses receive meat from more than one supplier. Again, different market shares are the reason for the geographical heterogeneity in the distribution of supply losses. Even though a sizable volume of pork meat is temporarily unavailable, the branching of the downstream network spreads this supply losses relatively uniformly across the population.

Figure 3 c) and Figure 3 d) show the supply levels of the population (like in Figure 2 b)). The x-axis shows the number of people that are affected, the y-axis the remaining supply levels in the respective districts. Note that the affected population sizes and the supply levels correspond to the coloring of districts in the upper panels a) and b). The horizontal line (dashed, red) indicates the chosen critical supply threshold of 0.66 (66%). The value on the x-axis where the supply level line intersects with the critical threshold indicates the number of people that suffer a supply reduction below the critical level. This value determines the systemic risk index, $SRI^{crit}$, for the respective business premise that failed. The volume-based systemic risk index, $SRI^{vol}$, can be read off the chart as the area (integral) between the supply level line (dark blue) and the constant horizontal line at 1 (100%)—indicated by the light blue shading—divided by the overall plotting area. It corresponds to the fraction of the supply that is affected overall in the country. In Figure 3 c) we see that a few districts within the capital fall to a supply level of around 40% and almost 2 million people are below the critical supply threshold of 66%.

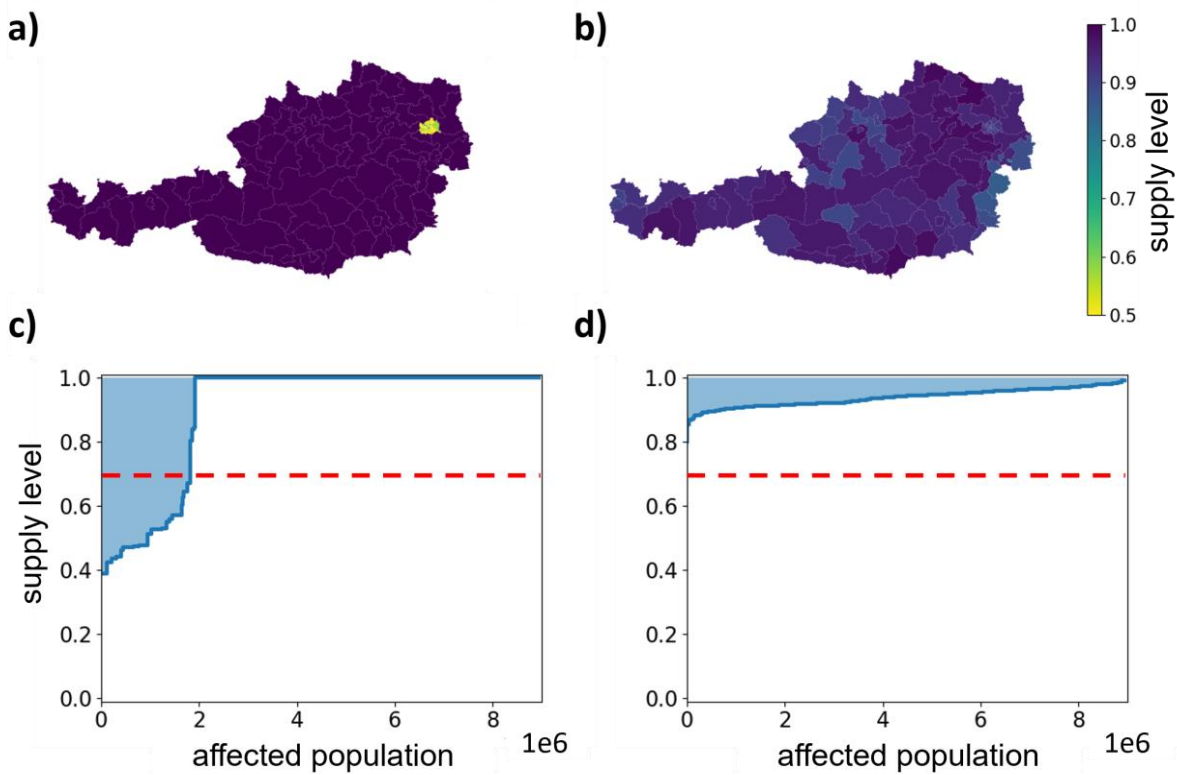

**Figure 3.** Visualization of simulated disruption propagation following the full disruption of operations of a single node in the SN.

The upper part—a) and b)—shows the remaining supply level of the population in each of the 116 districts. Light colors illustrate low, dark colors high supply levels. The lower part—c) and d)—is analogous to Figure 2 b), i.e., the x-axis is the affected population in the corresponding districts and the y-axis shows the remaining supply level of the respective district. Most affected districts (lowest remaining supply) are on the very left.

a) Estimated consequences for the pork meat supply of the population for the hypothetical failure of the riskiest distribution center. The effects are highly localized (affecting only the capital Vienna), but sizable (yellow color).

b) Simulated effects of the operational failure of a large slaughterhouse. The effects are more widespread than for the distribution center, but the respective districts are affected less severely (light blue colors).

c) For the simulated failure of the distribution center, the supply level drops below the critical threshold of 66% for approximately 20% of the population, implying a high systemic risk index value of $SRI^{crit} = 0.20$, and a comparably lower systemic risk index value, $SRI^{vol} = 0.10$.

d) For the simulated failure of the slaughterhouse the critical supply threshold is not surpassed for any district, the most affected district has a remaining supply level of below 90%. A majority of the districts suffer very small supply losses. $SRI^{crit} = 0$, and $SRI^{vol} = 0.06$.

The corresponding systemic risk indices are $SRI^{crit}$ = 0.20 and $SRI^{vol}$ = 0.10. Figure 3 d) shows that the critical supply threshold is not surpassed for any district, the most affected district has a remaining supply level of below 90%. A majority of the districts suffer very small supply losses. The values for the systemic risk indices are SRIcrit = 0.00  SRIvol = 0.06.

For the distribution center the volume based systemic risk index, SRIvol, is relatively low in comparison to the systemic risk index with respect to the critical supply threshold, SRIcrit, whereas the opposite is true for the slaughterhouse. This indicates that the volume-based systemic risk measure is by no means optimal for capturing the relevance of nodes in the SN when analyzing supply security. We advocate for using both measures to obtain a complete picture of the likely consequences of SN disruptions.

*Ranking of business premises according to their systemic risk index*

For becoming a meaningful tool for decision-makers the information contained in the simulation results must be reasonably aggregated without losing critical information. This implies a pre-selection of critical nodes for which an in-depth analysis could be useful as a basis for designing potential response strategies. $SRI^{crit}$ and $SRI^{vol}$, can be used as such aggregate measures that allow for a ranking to identify the most risky nodes in the SN. Next, we show the ranking for the Austrian pork SN.

The rankings are illustrated for the different node types separately—primary production, slaughterhouse, meat processor, distribution center and supermarket—to assess the role of the different node types. Figure 4 a) shows the ranking for the five different node types for the systemic risk index with respect to critical supply of the population, $SRI^{crit}$. The x-axis shows the rank of the node in each node category, the y-axis shows $SRI^{crit}$. Note that the x-axis is in log scale to shift the focus on the business premises with high $SRI^{crit}$ values. It is clearly visible that the distribution centers (red, dashed-dotted line) constitute the most risky category. The results suggest that 10 distribution centers can cause critical supply losses affecting more than 5% of the population and the riskiest affects around 20% of the population. Overall, approximately 20 distribution centers can cause critical losses. Slaughterhouses (orange dashed line) and meat processors (red dotted line) are the second and third most relevant categories, respectively. This result means that the riskiest slaughterhouse can affect around 7% of the population with a critical supply loss in pork meat. However, just five slaughterhouses can cause critical supply losses. Similarly, the most risky meat processor can cause critical supply losses of around 7% of the population and five meat processors are critical. Primary production nodes like farms, wholesalers, etc., (dark blue solid) and single supermarkets (purple dashed line) do not cause critical supply losses. The exception is one supermarket that critically affects one extremely small district with very few inhabitants only.

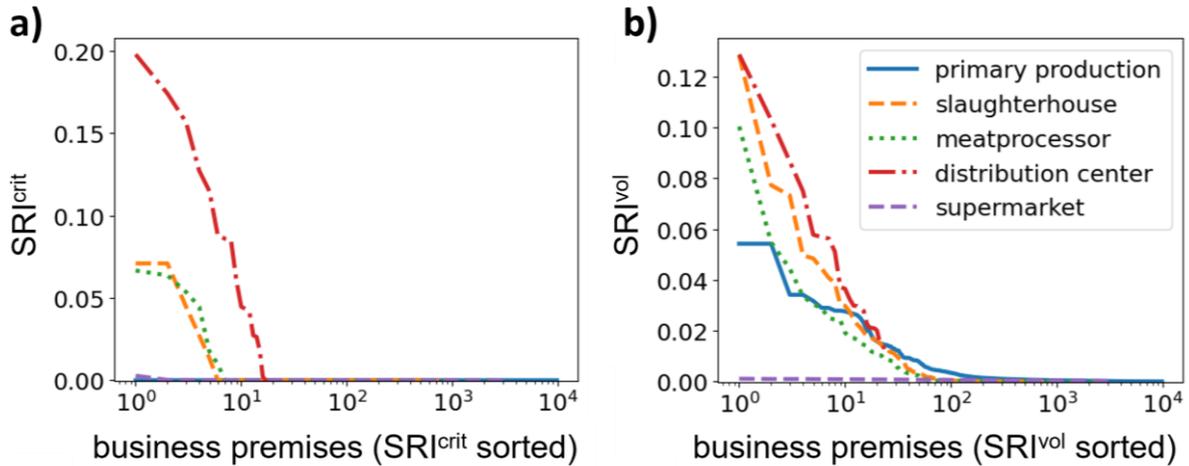

**Figure 4.** Systemic riskiness ranking for all 22,938 business premises in the pork supply network according to their node type.

a) Systemic risk index based on the critical supply threshold, $SRI^{crit}$, analogous to Figure 2 c). The x-axis (log-scale) shows the nodes ranked with respect to $SRI^{crit}$ for the five different node types. The most risky nodes are to the very left. Most premises do *not* cause a critical reduction of pork meat supply. The most critical premises are distribution centers (20 nodes) followed by slaughterhouses (5 nodes) and meat processors (5 nodes). No premise from the primary production (farms, wholesalers, etc.) are critical for the supply of the population.

b) Volume-based systemic risk index, $SRI^{vol}$. Distribution centers are the most risky type of premises in terms of overall volume, closely followed by slaughterhouses and meat processors. In terms of supply volumes also few primary production nodes can pose significant systemic risk.

Figure 4 b) shows the ranking of all business premises for the systemic risk index with respect to the overall supply volume of pork meat, SRIvol. The pattern is similar as in Figure 4 a) the distribution centers are most risky followed by the slaughterhouses and the meat processors. When looking at volumes only, it is apparent that also few primary production nodes have a sizable SRIvol of around 5%. This can be explained because the primary production category also includes wholesalers and live animal logistic premises that play an important role in supplying slaughterhouses with pigs and breeding farms supplying pigs to other large farms. These premises, however, do not cause critical supply losses, because of the downstream branching of the SN.

In future work the primary production sector could be decomposed into separate categories to provide a more detailed analysis. Single supermarkets are almost irrelevant with respect to SRIvol. For all node types it is visible that the decay of the systemic risk curves is slower than for SRIcrit in Figure 4 a), i.e., more nodes have a SRIvol visibly above zero. This difference occurs because for SRIvol every node that has a downstream connection to the population (districts) must have an SRIvol that is numerically greater than zero, even though being very small for most nodes. In contrast, SRIcrit identifies critical business premises. For non-critical business premises SRIcrit is zero, see Figure 4 a).

## DISCUSSION

Motivated by the Covid-19 pandemic and the aim to be better prepared for future crises, we address the following questions: "How do disruptions propagate through SNs of essential goods?", "How strong will the local population be affected by a disruption of a specific actor?", and "How can decision-makers identify those actors that carry high systemic risk, such that supply shortages in the population can be avoided, prepared for, or mitigated to ensure well-being?" We design a practical assessment tool that would allow governments and decision-makers to assess systemic risks in SN. In particular, we develop a rigorous methodology to simulate expected local supply losses (in each district) caused by a specific SN disruption scenario. We suggest a systemic risk index, $SRI^{crit}$, that serves to rank business premises with respect to their relevance for the supply of a population.

Results show that business premises with high volume and strong geographical concentration of supply losses are the most critical. Those with large supply volumes and a widespread geographical impact are not critical, because they do affect a large part of the population, however, not critically. Business premises with smaller supply volumes but a highly localized impact can critically affect small parts of the population but are not critical on a wider geographical scale. Low volume and low concentration nodes do not require systemic risk management actions. Findings are summarized in Figure 5.

We empirically demonstrate the risk assessment tool based on a nation-wide pork SN that includes all relevant actors. Results show that for the nation's pork supply only very few nodes are likely to cause critical supply shortages in the basic supply of the population. These are readily identified with the suggested systemic risk index, $SRI^{crit}$. The methodology further allows for a more in-depth analysis of the distribution of supply losses across local populations in districts and geography. Findings confirm the intuitive expectation that distribution centers of supermarket chains are highly critical, but also other types of business premises can be critical nodes. Institutions responsible for food supply security should be aware of these nodes, for effective crisis management and for planning support strategies before and during crises scenarios. In summary, our study significantly expands supply chain management literature and at the same time it is relevant to practitioners including governments, civil defense/protection, and other decision makers, since the simulation environment enables them to better protect the population from disruption induced supply shortages of essential goods.

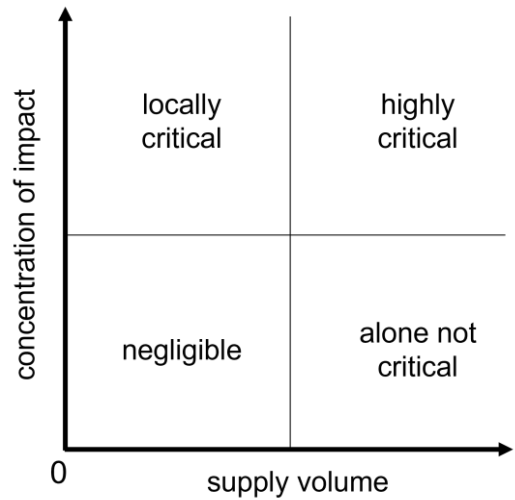

**Figure 5.** Risk factors of business premises for a population's basic supply. The x-axis denotes the volume of supplied goods from the premise, the y-axis denotes geographical concentration of impacts on the population.

**Contribution to theory**

This study has several implications on the supply chain management literature. First, much of the existing literature focuses on one focal firm or dyadic buyer-supplier relationships as the unit of analysis. To date, only few have examined supply chain *networks*. To date the understanding of SNs and their impact on firms, and on society more generally, is limited. Previous empirical studies examined SNs from the perspective of a focal firm, with a strong focus on the automotive sector (e.g., Choi & Hong, 2001; Kim *et al.,* 2011; Shao *et al.,* 2018; Potter & Wilhelm, 2020). Others capture a small extract of other industries' SNs, such as a focal firm's first (or second) tier supply base (Lu & Shang, 2017; Dong *et al.,* 2020). Collecting data on entire SNs can be demanding. Often firms do not disclose their relationships; even if data collection is technically possible the related costs can be enormous. Two recent network studies, i.e., Wiedmer & Griffis (2021) and Gualandris *et al.* (2021), investigate five-tier SNs of 21 firms from the Mergent Horizon® database, and 4803 firms and 20,504 contractual ties organized in 187 extended supply chains from Bloomberg SPLC and ESG, respectively. Even fewer studies examine more general SNs that are not collected via a snowball-scheme-like sampling approach starting from single firms but aim to analyze the entirety of the SN within predefined boundaries (Borsos & Stancsics, 2020; Dhyne *et al.,* 2021; Diem *et al.,* 2021). We build on and contribute to these studies by examining a large-tiered, nation-wide pork supply network, including all relevant actors from primary production to final consumers. We start to bridge the gap from a focal firm/dyadic perspective to a truly supply chain network perspective.

Second, supply chain management literature has examined risks imposed by supply chain disruptions on a variety of firm performance outcomes, related to operations, finance, and relationships (Hendricks & Singhal, 2005; Craighead *et al.,* 2007; Wu *et al.,* 2007; Hendricks *et al.,* 2009; Bode & Macdonald, 2017). An emerging stream of research on transformative supply chain research (TSCR) (Mollenkopf *et al.,* 2020), emphasizes the importance of social outcomes and highlights the critical, yet overlooked role of supply chains on the well-being of individuals and communities (Mollenkopf *et al.,* 2021). Transformative consumer research (TCR) in marketing and transformative service research (TSR) in management already examine topics such as poverty alleviation (Blocker et al., 2013), sustainability (Prothero *et al.,* 2011), and health (Anderson *et al.,* 2017), aiming to improve well-being. Yet, the TSCR literature is still limited. This present study contributes to TSCR by developing a novel methodology that allows us to assess the vulnerability of the population to critical supply shortages caused by major disruptions. We show how TSCR can support governments and decision-makers in ensuring the supply of the population with essential goods in times of crisis.

Third, we contribute to the systemic risk literature. Systemic risk, even though of growing concern, is difficult to measure and quantify (Montagna *et al.,* 2020). So far, systemic risk has been predominantly studied in the financial literature (e.g., Boss, *et al.*, 2004; Haldane & May, 2011; Battiston *et al.,* 2012; Thurner & Poledna, 2013; Poledna & Thurner, 2016; Glasserman & Young, 2016; Diem, *et al.*, 2020; Bardoscia *et al.,* 2021). However, understanding systemic risk is also essential to maintaining SN stability (Scheibe & Blackhurst, 2018). Despite its importance, SCM literature on systemic risks is scarce, with a notable, yet conceptual, exception of Scheibe & Blackhurst (2018), who employed a grounded theory approach to better understand the drivers of supply chain disruption propagation. We contribute to this research by developing a risk index that allows a ranking of firms in large scale SNs according to their systemic riskiness for the population. We apply it to the level of individual business premises opposed to the regular firm-level analysis. An important aspect that differentiates our study from many others is our disruption propagation algorithm. It is based on continuous state variables measuring the exact disruption level of a firm's operations—in comparison to the binary state variables that are commonly used—and it explicitly includes the supply levels of the local population. As such, it fosters an explicitly population-focused systemic risk index. We showcase its practical usefulness on the unique business premises level SN of the nation-wide pork production of Austria.

**Practical implications**

The presented risk assessment tool has immediate practical implications for the management of crises and disasters—such as the current Covid-19 pandemic—that cause severe disruptions in SNs. The suggested systemic risk index quantifies the riskiness of all actors in a given supply network and identifies firms and business premises that can potentially cause major disruptions. Policy makers can engage highly risky firms in crisis management boards to build up direct communication channels and foster coordinated crisis responses. The Covid-19 pandemic demonstrated the importance of coordinated responses at all levels in order to avoid adverse effects of the welfare of the population (Haug *et al.,* 2020; Li, Chandra, & Kapucu, 2020). When transparently communicated to firms ranked critical, they become aware of their position in the SNs and their status for others. They could develop supply chain resilience capabilities that enable them to withstand or quickly recover from disruptions (Polyviou *et al.,* 2020).

For companies, this study highlights the value of supply chain trust. Extant research has already shown that trust improves firm performance (Cao & Zhang, 2011; Fawcett *et al.*, 2012), however, organizational conflict and fear of opportunistic behavior still hinders firms to share sensitive information (Fawcett *et al.*, 2017). This research project is unique in that firms within a whole SN—including competitors—share sensitive data. We show that trust to share data for the purpose of ensuring population well-being has also immediate advantages for the individual companies. That is, companies receive real-time information of critical events within the SN. Too often companies focus only on strategic, top-tier suppliers (Yan *et al.*, 2015). Yet, a critical supplier "can come from anywhere in a multitier supply network" (Yan *et al.*, 2015, p. 52). When a disruptive event happens several tiers upstream in the SN, it is likely to remain hidden to the potentially affected business premises for quite a time—delaying response time and increasing the impact as the disruption cascades through the network. Though, if companies are alert early, they can engage in countermeasures in a timely and effective manner—avoiding severe impact.

Finally, the developed methodology is generalizable. Accordingly, policy makers can simulate the effects of any specific stress scenario and draft countermeasures against their most likely effects. This study shows that if researchers, practitioners, and governments collaborate and trust each other, we are not only able to advance theory and methodology but achieve actual impact by demonstrating how supply chain management research can contribute to effective crisis preparedness and thus supports firm performance as well as individual and community well-being.

**Limitations and future research**

Quantitative studies based on real data often suffer limitations due to model assumptions and data availability. Although we demonstrated our developed tool to work on a data set that covers almost the entire nation-wide pork supply chain, there exist parts of the network for which actual flow data could not yet be collected. There we had to estimate these missing data (see Appendix C for details). Due to data unavailability, we can at this point not include supporting materials, like packaging in the production process, in our model. Yet, suppliers of supporting materials potentially carry significant systemic risk. For example, a single packaging manufacturer supplying several large production plants might be highly critical for the overall system. The algorithm is able to include these actors, as soon as the respective data becomes available. Future work is necessary to extend the methodology to a multi-product SN, where all major food and supporting supplies are integrated.

Moreover, a number of further analyses can be conducted with the presented methodology. For example, for each district one could calculate how many nodes pose a critical threat to the district and one could define a resilience index for every district. In this present study rather coarse grained administrative districts were chosen for the analysis. Future extensions could include estimations of driving times from smaller population nodes represented by more fine-grained administrative districts (for example electoral districts) to supermarkets. This would take the fact into account that people will drive to other districts when local supply depletes. Another future step would be to include inventory levels of supply network premises. Finally, in developing countries the presented methodology could have an impact on the exposure of the population to draughts, floods, and other catastrophes affecting primary production. From a transformative research perspective, increased efforts in building similar data sets for the most vulnerable parts of the global population would be essential.

**CONCLUSION**

We examine supply disruption propagation effects and develop a systemic risk assessment tool that enables decision-makers to identify systemically critical actors, quantify the impact of their hypothetical default, and assess how a supply disruption would propagate through the network and how it will, finally, affect the population locally. This study is among the few in SCM to investigate (the risk of) supply disruption propagation to examine the impact of disruptions on the well-being of individuals and communities. We demonstrate the applicability of the risk assessment tool on a nation-wide food (pork) supply network of approximately 23,000 business premises, 44,000 supply links and 116 administrative districts. The study is among the few that examine weighted and directed large scale real-world SNs and findings expand and enrich SCM theory, as well as methodology of quantifying risk propagation in SNs. Finally, our work contributes to transformative research in that we not only develop a rigorous methodology, but also disseminate it to relevant stakeholders to improve human well-being during crises.

# REFERENCES

Albert, R., Jeong, H., & Barabási, A. L. (2000). Error and attack tolerance of complex networks. *Nature*, 406(6794), 378-382.

Anderson, L. & Ostrom, A. (2015). Transformative service research: Advancing our knowledge about service and well-being, *Journal of Service Research*, 18(3), 243-249.

Autry, C. W., & Griffis, S. E. (2008). Supply chain capital: the impact of structural and relational linkages on firm execution and innovation. *Journal of Business Logistics*, 29(1), 157-173.

Austrian Ministry of Agriculture (2021) "Grüner Bericht 2021 — Die Situation der österreichischen Land- und Forstwirtschaft Gemäß §9 des Landwirtschaftgesetzes"

Bardoscia, M., Barucca, P., Battiston, S., Caccioli, F., Cimini, G., Garlaschelli, D., Saracco, F., Squartini, T., & Caldarelli, G. (2021). The physics of financial networks. *Nature Reviews Physics*, 1-18.

Basole, R. C., & Bellamy, M. A. (2014). Supply network structure, visibility, and risk diffusion: A computational approach. *Decision Sciences*, 45(4), 753-789.

Battiston, S., Puliga, M., Kaushik, R., Tasca, P., & Caldarelli, G. (2012). Debtrank: Too central to fail? financial networks, the fed and systemic risk. Scientific Reports, 2(1), 1-6.

Blocker, C.P., Ruth, J.A., Sridharan, S., Beckwith, C., Ekici, A., Goudie-Hutton, M., Rosa, J., Saatcioglu, B., Trujillo, C., Talukdar, D., and Varman, R. (2013). Understanding Poverty and Promoting Poverty Alleviation through Transformative Consumer Research, *Journal of Business Research*, 66(8), 1195-1202.

Bode, C., & Macdonald, J. R. (2017). Stages of supply chain disruption response: Direct, constraining, and mediating factors for impact mitigation. *Decision Sciences*, 48(5), 836-874.

Bode, C., & Wagner, S. M. (2015). Structural drivers of upstream supply chain complexity and the frequency of supply chain disruptions. *Journal of Operations Management*, 36, 215-228.

Bode, C., Macdonald, J. R., & Merath, M. (2021). Supply disruptions and protection motivation: Why some managers act proactively (and others don't). *Journal of Business Logistics*, 00, 1-24.

Bode, C., Wagner, S. M., Petersen, K. J., & Ellram, L. M. (2011). Understanding responses to supply chain disruptions: Insights from information processing and resource dependence perspectives. *Academy of Management Journal*, 54(4), 833-856.

Borgatti, S. P., & Foster, P. C. (2003). The network paradigm in organizational research: A review and typology. *Journal of Management*, 29(6), 991-1013.

Borsos, A., & Stancsics, M. (2020). Unfolding the hidden structure of the Hungarian multi-layer firm network. Central Bank of Hungary Working Paper (No. 139).

Boss, M., Elsinger, H., Summer, M., & Thurner 4, S. (2004). Network topology of the interbank market. *Quantitative finance*, *4*(6), 677-684.

Boss, M., Summer, M., & Thurner, S. (2004). Contagion Flow through Banking Networks. In: *International Conference on Computational Science*. Springer. 2004, pp. 1070–1077.

Brass, D. J., Galaskiewicz, J., Greve, H. R., & Tsai, W. (2004). Taking stock of networks and organizations: A multilevel perspective. *Academy of Management Journal*, 47(6), 795-817.

Burgos, D., & Ivanov, D. (2021). Food retail supply chain resilience and the COVID-19 pandemic: A digital twin-based impact analysis and improvement directions. *Transportation Research Part E: Logistics and Transportation Review*, 152, 102412.

Burkacky, O., Lingemann, S., & Pototzky, P. (2021). Coping with the auto-semiconductor shortage: Strategies for success. WWW: https://www.mckinsey.com/industries/automotive-and-assembly/our-insights/coping-with-the-auto-semiconductor-shortage-strategies-for-success [accessed 12/28/2021].

Cao, M., & Zhang, Q. (2011). Supply chain collaboration: Impact on collaborative advantage and firm performance. *Journal of Operations Management*, *29*(3), 163-180.

Choi, T. Y., & Hong, Y. (2002). Unveiling the structure of supply networks: case studies in Honda, Acura, and DaimlerChrysler. *Journal of Operations Management*, 20(5), 469-493.

Choi, T. Y., Dooley, K. J., & Rungtusanatham, M. (2001). Supply networks and complex adaptive systems: control versus emergence. *Journal of Operations Management*, 19(3), 351-366.

Choi, T. Y., Narayanan, S., Novak, D., Olhager, J., Sheu, J. B., & Wiengarten, F. (2021). Managing extended supply chains. *Journal of Business Logistics*, 42, 200-206.

Christopher, M., Peck, H. (2004). Building the Resilient Supply Chain", *The International Journal of Logistics Management*, 15(2), 1-14.

Cont, R., & Schaanning, E. (2019). Monitoring indirect contagion. *Journal of Banking & Finance*, 104, 85-102.

Cont, R., Moussa, A., & Santos, E. (2010). Network Structure and Systemic Risk in Banking Systems. *SSRN.* 1733528. https://dx.doi.org/10.2139/ssrn.1733528

Cont, R., Moussa, A., & Santos, E. (2013). Network Structure and Systemic Risk in Banking Systems. In J. Fouque & J. Langsam (Eds.), Handbook on Systemic Risk (pp. 327-368). Cambridge: Cambridge University Press.

Craighead, C. W., Blackhurst, J., Rungtusanatham, M. J., & Handfield, R. B. (2007). The severity of supply chain disruptions: design characteristics and mitigation capabilities. *Decision Sciences*, 38(1), 131-156.

Craighead, C. W., Ketchen Jr, D. J., & Darby, J. L. (2020). Pandemics and supply chain management research: toward a theoretical toolbox. *Decision Sciences*, 51(4), 838-866.

Crucitti, P., Latora, V., & Marchiori, M. (2004). Model for cascading failures in complex networks. *Physical Review E*, 69(4), 045104.

Dhyne, E., Kikkawa, A. K., Mogstad, M., & Tintelnot, F. (2021). Trade and domestic production networks. *The Review of Economic Studies*, *88*(2), 643-668.

Diem, C., Borsos, A., Reisch, T., Kertész, J., & Thurner, S. (2021). Quantifying firm-level economic systemic risk from nation-wide supply networks. Available at SSRN 3826514.

Diem, C., Pichler, A., & Thurner, S. (2020). What is the minimal systemic risk in financial exposure networks?. *Journal of Economic Dynamics and Control*, *116*, 103900.

Dolgui, A., Ivanov, D. & Sokolov, B. (2018) Ripple effect in the supply chain: an analysis and recent literature, *International Journal of Production Research*, 56:1-2, 414-430.

Dong, Y., Skowronski, K., Song, S., Venkataraman, S., & Zou, F. (2020). Supply base innovation and firm financial performance. *Journal of Operations Management*, 66(7-8), 768-796.

Eisenberg, L., & Noe, T. H. (2001). Systemic Risk in Financial Systems. *Management Science*, 47(2), 236-249.

European Commission (2021). Contingency plan for ensuring food supply and food security in times of crisis. COM/2021/689 final.

Falcone, E., Fugate, B., & Dobrzykowski, D. (2021). Supply chain plasticity during a global disruption: Effects of CEO and supply chain networks on operational repurposing. *Journal of Business Logistics*.

Fawcett, S. E., Fawcett, A. M., Watson, B. J., & Magnan, G. M. (2012). Peeking inside the black box: toward an understanding of supply chain collaboration dynamics. *Journal of Supply Chain Management*, *48*(1), 44-72.

Fawcett, S. E., Jin, Y. H., Fawcett, A. M., & Magnan, G. (2017). I know it when I see it: the nature of trust, trustworthiness signals, and strategic trust construction. *The International Journal of Logistics Management*.

Fujiwara, Y., Terai, M., Fujita, Y., & Souma, W. (2016). Debtrank analysis of financial distress propagation on a production network in Japan. RIETI Discussion Paper Series 16-E-046.

Gerschberger, M., Engelhardt-Nowitzki, C., Kummer, S., & Staberhofer, F. (2012). A model to determine complexity in supply networks. *Journal of Manufacturing Technology Management*. 23(8), 1015-1037.

Glasserman, P. & Young, H. (2016). Contagion in financial networks. *Journal of Economic Literature*, 54(3), 779-831.

Gualandris, J., Longoni, A., Luzzini, D. & Pagell, M. (2021). The association between supply chain structure and transparency: A large-scale empirical study. *Journal of Operations Management*, 67(7), 803-827.

Haldane, A. G., & May, R. M. (2011). Systemic risk in banking ecosystems. *Nature*, 469(7330), 351-355.

Haug, N., Geyrhofer, L., Londei, A., Dervic, E., Desvars-Larrive, A., Loreto, V., Pinior, B., Thurner, S., & Klimek, P. (2020). Ranking the effectiveness of worldwide COVID-19 government interventions. *Nature Human Behaviour* 4, 1303–1312.

Hendricks, K. B., & Singhal, V. R. (2005). An empirical analysis of the effect of supply chain disruptions on long-run stock price performance and equity risk of the firm. Production and Operations Management, 14(1), 35-52.

Hendricks, K. B., Singhal, V. R., & Zhang, R. (2009). The effect of operational slack, diversification, and vertical relatedness on the stock market reaction to supply chain disruptions. *Journal of Operations Management*, 27(3), 233-246.

Hsiao, H. I., Kemp, R. G. M., Van der Vorst, J. G. A. J., & Omta, S. O. (2010). A classification of logistic outsourcing levels and their impact on service performance: Evidence from the food processing industry. *International Journal of Production Economics*, 124(1), 75-86.

Inoue, H., & Todo, Y. (2019). Firm-level propagation of shocks through supply-chain networks. *Nature Sustainability*, *2*(9), 841-847.

Ivanov, D., & Dolgui, A. (2020). Viability of intertwined supply networks: extending the supply chain resilience angles towards survivability. A position paper motivated by COVID-19 outbreak. *International Journal of Production Research*, 58(10), 2904-2915.

Ketchen Jr, D. J., & Craighead, C. W. (2020). Research at the intersection of entrepreneurship, supply chain management, and strategic management: opportunities highlighted by COVID-19. *Journal of Management*, 46(8), 1330-1341.

Kim, Y., Chen, Y. S., & Linderman, K. (2015). Supply network disruption and resilience: A network structural perspective. *Journal of Operations Management*, 33, 43-59.

Kim, Y., Choi, T. Y., Yan, T., & Dooley, K. (2011). Structural investigation of supply networks: A social network analysis approach. *Journal of Operations Management*, 29(3), 194-211.

Knoke, D., & Yang, S. (2019). Social network analysis. Third Edition. SAGE Publications, Inc . https://doi.org/10.4135/97815 06389332

Ledwoch, A., Brintrup, A., Mehnen, J., & Tiwari, A. (2016). Systemic risk assessment in complex supply networks. *IEEE Systems Journal*, *12*(2), 1826-1837.

Li, Y., & Zobel, C. W. (2020). Exploring supply chain network resilience in the presence of the ripple effect. *International Journal of Production Economics*, 228, 107693.

Li, Y., Chandra, Y., & Kapucu, N. (2020). Crisis coordination and the role of social media in response to COVID-19 in Wuhan, China. *The American Review of Public Administration*, 50(6-7), 698-705.

Li, Y., Chen, K., Collignon, S., & Ivanov, D. (2021). Ripple effect in the supply chain network: Forward and backward disruption propagation, network health and firm vulnerability. *European Journal of Operational Research*, 291(3), 1117-1131.

Lu, G., & Shang, G. (2017). Impact of supply base structural complexity on financial performance: Roles of visible and not-so-visible characteristics. *Journal of Operations Management*, 53, 23-44.

Mollenkopf, D. A., Ozanne, L. K., & Stolze, H. J. (2020). A transformative supply chain response to COVID-19. *Journal of Service Management*.

Mollenkopf, D., Stolze, H., Esper, T., Ozanne, L. (2021). Special Topic Forum Call for Papers: Transformative Supply Chain Research. *Journal of Business Logistics*.

Montagna, M., Torri, G., & Covi, G. (2020). On the origin of systemic risk. Available at SSRN 3699369.

Nair, A., & Vidal, J. M. (2011). Supply network topology and robustness against disruptions–an investigation using multi-agent model. *International Journal of Production Research*, 49(5), 1391-1404.

Perez, C., de Castro, R., Simons, D., & Gimenez, G. (2010). Development of lean supply chains: a case study of the Catalan pork sector. *Supply Chain Management: An International Journal*.

Pettit, T. J., Croxton, K. L., & Fiksel, J. (2013). Ensuring supply chain resilience: development and implementation of an assessment tool. *Journal of Business Logistics*, 34(1), 46-76.

Pinior, B., Conraths, F., Petersen, B., & Selhorst, T. (2015). Decision support for risks managers in the case of deliberate food contamination: The dairy industry as an example. *Omega*, 53, 41-48.

Pinior, B., Konschake, M., Platz, U., Thiele, D. H., Petersen, B., Conraths, F., & Selhorst, T. (2012b). The trade network in the dairy industry and its implication for the spread of contamination. *Journal on Dairy Science*, 95(11), 6351-6361.

Pinior, B., Platz, U., Ahrens, U., Petersen, B., Conraths, F., & Selhorst, T. (2012a). The German milky way: trade structure of the milk industry and possible consequences of a food crisis. *Journal on Chain and Network Science*, 12(1), 25-39.

Poledna, S., & Thurner, S. (2016). Elimination of systemic risk in financial networks by means of a systemic risk transaction tax. *Quantitative finance*, *16*(10), 1599-1613.

Polyviou, M., Croxton, K. L., & Knemeyer, A. M. (2020). Resilience of medium-sized firms to supply chain disruptions: the role of internal social capital. *International Journal of Operations & Production Management*, 40(1), 68-91.

Potter, A., & Wilhelm, M. (2020). Exploring supplier–supplier innovations within the Toyota supply network: A supply network perspective. *Journal of Operations Management*, 66(7-8), 797-819.

Prothero, A., Dobscha, S., Freund, J., Kilbourne, W.E., Luchs, M.G., Ozanne, L.K. and Thøgersen, J. (2011). Sustainable consumption: Opportunities for consumer research and public policy, *Journal of Public Policy & Marketing*, 30(1), 31–38.

Reisch, T., Heiler, G., Diem, C., & Thurner, S. (2021). Inferring supply networks from mobile phone data to estimate the resilience of a national economy. *arXiv preprint arXiv:2110.05625*.

Scheibe, & Blackhurst, J. (2019). Systemic Risk and the Ripple Effect in the Supply Chain. In *Handbook of Ripple Effects in the Supply Chain* (pp. 85–100). Springer International Publishing.

Scheibe, K. P., & Blackhurst, J. (2018). Supply chain disruption propagation: a systemic risk and normal accident theory perspective. *International Journal of Production Research*, 56(1-2), 43-59.

Shao, B. B., Shi, Z. M., Choi, T. Y., & Chae, S. (2018). A data-analytics approach to identifying hidden critical suppliers in supply networks: Development of nexus supplier index. *Decision Support Systems*, 114, 37-48.

Sharma, A., Pathak, S., Borah, S. B., & Adhikary, A. (2020). Is it too complex? The curious case of supply network complexity and focal firm innovation. *Journal of Operations Management*, 66(7-8), 839-865.

Sheffi, Y., & Rice Jr, J. B. (2005). A supply chain view of the resilient enterprise. *MIT Sloan Management Review*, 47(1), 41.

Taylor, D. H. (2006). Strategic considerations in the development of lean agri-food supply chains: a case study of the UK pork sector. *Supply Chain Management: An International Journal*, 11/3, 271-280.

Thurner, S., & Poledna, S. (2013). DebtRank-transparency: Controlling systemic risk in financial networks. *Scientific reports*, *3*(1), 1-7.

Van Hoek, R. (2020). Research opportunities for a more resilient post-COVID-19 supply chain–closing the gap between research findings and industry practice. *International Journal of Operations & Production Management*, 40(4), 341-355.

Vanpoucke, E., & Ellis, S. C. (2020). Building supply-side resilience–a behavioural view. *International Journal of Operations & Production Management*.

Vlajic, J. V., Van der Vorst, J. G., & Haijema, R. (2012). A framework for designing robust food supply chains. *International Journal of Production Economics*, 137(1), 176-189.

Wenzel, M., Stanske, S., & Lieberman, M. B. (2020). Strategic responses to crisis. *Strategic Management Journal*, 41(7/18).

Wiedmer, R., & Griffis, S. E. (2021). Structural characteristics of complex supply chain networks. *Journal of Business Logistics*, 42(2), 264-290.

Wiedmer, R., Rogers, Z. S., Polyviou, M., Mena, C., & Chae, S. (2021). The dark and bright sides of complexity: A dual perspective on supply network resilience. *Journal of Business Logistics*, 42(3), 336-359.

Wu, T., Blackhurst, J., & O'Grady, P. (2007). Methodology for supply chain disruption analysis. *International Journal of Production Research*, 45(7), 1665-1682.

Yan, T., Choi, T. Y., Kim, Y., & Yang, Y. (2015). A theory of the nexus supplier: A critical supplier from a network perspective. *Journal of Supply Chain Management*, *51*(1), 52-66.

## APPENDIX

**Appendix A: Notation for networks and network measures**

This section builds the necessary mathematical notation to define the disruption propagation algorithm and to define the systemic risk indices SRI^crit and SRI^vol. We work with a weighted directed supply network with $n$ nodes, representing the different business premises of firms and $m$ nodes, representing the administrative districts where the population is located and that are considered for the empirical analysis. The network is represented by the weighted adjacency matrix $W$, of dimension $(n+m) \times (n+m)$. The nodes $i \in \{1,2,\ldots,n\}$ are supply nodes and the nodes $k \in \{n+1, n+2, \ldots, n+m\}$. The element $W_{ij}$ represents the volume of products supplied from node $i$ to node $j$ in a given period of time. The vector, $p$, represents the node type for all $n$ nodes, i.e., the element $p_i$ represents the category type of business premise $i$. We consider five different types of supply network nodes namely primary production, slaughterhouse, meat processor, distribution center, supermarket and population. The category primary production is a summary term for different types of farms (e.g., breeding or fattening) and service facilities like wholesale traders, and logistic providers. We assume that the type of products delivered between node types is determined by the node types. Primary production nodes deliver living pigs to other primary production nodes and slaughterhouses. Slaughterhouses deliver pig carcasses to meat processors and cut meat parts to distribution centers directly. Meat processors deliver processed meat to distribution centers. Distribution centers deliver the received types of meat proportionally to their supermarkets and the local supermarkets within a district supply the local population of this district. The number of inhabitants living in district $k$ is denoted as $e_k$ and the total number of inhabitants is $e^{tot} = \sum_{k=1}^{m} e_k$.

From the weighted adjacency matrix, $W$, we calculate the in-strength of node $i$ as $s_i^{in} = \sum_{j=1}^{n} W_{ji}$ and the out-strength of node $i$ as $s_i^{out} = \sum_{j=1}^{n} W_{ji}$. For each node $i$ we define the regular absolute production level when the supply network, $W$, is undisrupted as $x_i(0) = \sum_{j=1}^{n} W_{ji}$. The absolute production level of the node $i$ at iteration $t$ of the disruption propagation algorithm is denoted as $x_i(t) = \sum_{j=1}^{n} W_{ji} h_j(t)$. Where for each node $i$ in the network we define the second state variable, $h_i(t) \in [0,1]$, representing the relative amount of production of node $i$ at iteration $t$, i.e., $h_i(t) = 1$ means 100% production and 0% disruption while $h_i(t) = 0$ means no production and 100% disruption. For the undisrupted supply network, we initially have $h_i(0) = 1$ for all nodes $i$. Note the relation $h_i(t) = \frac{x_i(t)}{x_i(0)}$.

For each population node $k$ we define the indicator variable $\delta_k$ indicating whether the supply level in a population node (district) fell below the critical threshold $\lambda$, i.e., in mathematical notation, $\delta_k = 1$ if $h_k(T) < \lambda$ and $\delta_k = 0$ if $h_k(T) \geq \lambda$. In the empirical analysis we set $\lambda = 0.66$. Now we can define the systemic risk index with respect to the critical supply level of the population for node $i$ as

$$SRI_i^{crit} = \sum_{k=n+1}^{n+m} \frac{e_k}{e^{tot}} (1 - h_k(T)) \delta_k \qquad \text{(Eq. 1)}$$

The systemic risk index with respect to to the overall supply volume reduction is defined for node $i$ as

$$SRI_i^{vol} = \sum_{k=n+1}^{n+m} \frac{e_k}{e^{tot}} (1 - h_k(T)) \qquad \text{(Eq. 2)}$$

Note the differences to the definition of the ESRI in Diem *et al.* (2021) where no critical threshold is used and the supply of the population is not considered, but the overall production level in the network.

For a population node $k$ the weighted in-link, $W_{ik}$, from a supermarket $i$ represents the volume of supply the district receives from the supermarket. Thus, the in-strength, $s_i^{in}$, of a population node $k$ corresponds to its total supply volume when the supply network $W$ is undisrupted. Therefore, for the population node $k$ the value of $h_k(t)$ correctly denotes the available supply in the district $k$ at iteration $t$.

# Appendix B: Simulation methodology for propagation of disruptions

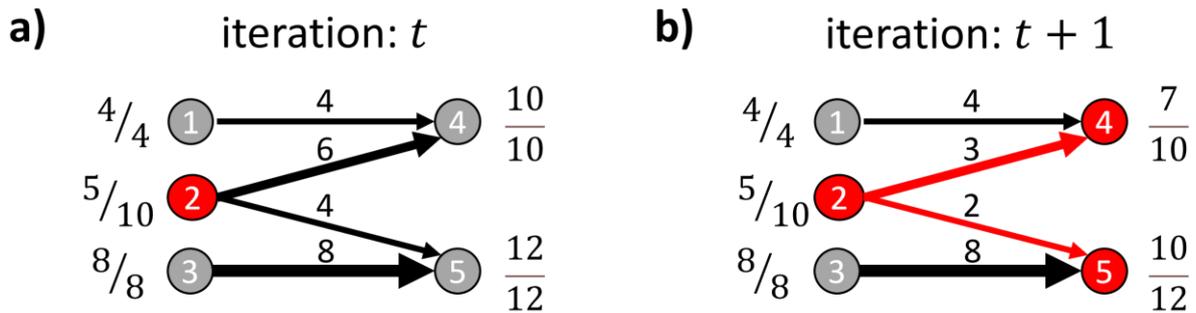

**Figure A.1.** Example of one iteration step in the disruption propagation simulation. The fractions next to the nodes indicate the production level at the respective iteration step relative to the undisrupted production level. At iteration t node 2 is affected by a disruption (stemming from its suppliers omitted in this illustration) and the production level of node 2 drops from 10 to 5 (i.e., the relative production level is $h_2(t) = 0.5$). At time t+1 the disruption propagates to the buyers of node 2 and their production level drops from 10 to 7 and 12 to 10 for nodes 4 and 5, respectively, i.e., the relative production levels are $h_4(t+1) = 0.7$ and $h_5(t+1) = 0.8333$. In the next iteration nodes 4 and 5 will forward their disruption downstream to their own buyers (omitted in the illustration).

Now, we describe intuitively how the downstream disruption propagation is simulated in our model. As explained in the data section we focus on the application of one specific product type due to the data availability. The model can be readily extended to a multi-product type supply network after collecting data on other supply chains.

First, we introduce the variable, $h_i(t)$, that captures how severely a node $i$ is disrupted over the course of the simulation. It can range between 0 – meaning fully disrupted – and 1 – meaning fully functional. For each node the state variable can be interpreted as its current production level in relation to its undisrupted production level and is simply calculated as the output the node can currently deliver to its customer divided by the amount it could deliver initially before any disruption occurred in the network.

Similarly, we have to devise a state variable, $h_k(t)$, for each population node $k$ that corresponds to the access to basic goods sold in supermarkets. The variable ranges between 0, no access to essential supply and 1, full access to the essential supply (like in the undisrupted network). It is calculated as the current amount of basic supply products available at the supermarkets within the district divided by the initial amount of basic supply products available in those supermarkets.

Before the start of the disruption propagation, we assume an initially stable state where the state variable of all supply network nodes and population nodes are 1, meaning no disruption occurred. Then, in the first iteration of the simulation we set the state variable of the node that is assumed to be initially disrupted to 0, i.e., it is fully disrupted and cannot deliver any products to its buyers. In the second iteration the buyers of the initially disrupted node, i.e., nodes having a downstream connection to the disrupted node, are affected by the disruption of their supplier. The disruption is assumed to be proportional to the importance of the supplier for the respective buyer and the respective product type. This is simulated by calculating the fraction of lost supply of each buying node, by putting the inflows stemming from the currently disrupted node into proportion to the buying node's overall inflows (of the respective product types). In mathematical terms this is simply

$$h_i(t+1) = \sum_{j=1}^{n} \frac{W_{ji}}{s_i^{in}} h_j(t) \quad .$$ (Eq. 3)

Next, the lost fraction of inflows at the buying node leads to a decrease in the product outflows of this node, which in turn will again affect the buyers of this node. This continues until no new disruptions occur.

Figure 3 illustrates how a disrupted node propagates its disruption downstream to its buyers. The fractions next to the nodes indicate the production level at the respective iteration step relative to the undisrupted production level. The numerator indicates the current level of available output and the denominator the initial level of available output. At iteration $t$ *node 2* is affected by a disruption (stemming from its suppliers omitted in this illustration) and the production level of *node 2* drops from 10 to 5 (i.e., the relative production level is $h_2(t) = 0.5$). At time $t+1$ the disruption propagates to the buyers of *node 2* and their production level drops from 10 to 7 and 12 to 10 for *nodes 4* and *5*, respectively, i.e., $h_4(t+1) = 0.7$ and $h_5(t+1) = 0.833$. In the next iteration *nodes 4* and *5* will forward their disruption downstream to their own buyers (omitted in the illustration).

These dynamics can be mathematically compactly defined as a recursion algorithm that allows to estimate the supply level of each population node (district) depending on the operations of all other indirectly connected upstream nodes, like in Eq. 3. Note that the upstream propagation can be defined analogously. Up- and downstream propagation can also be treated differently when more data about the production process is available (Diem et. al, 2021). We summarize the simulation of downstream disruption propagation in Algorithm 1.

**Algorithm 1.** Simulation of downstream disruption propagation.

1. **initialize** each node with production level 100%, i.e., $h_j(0) = 1$, and set production of the initially disrupted node $i \neq j$ to 0%, $h_i(1) = 0$.
2. **while** production levels of nodes are lower than in previous iteration **do**
    a. **for** each node with lower than previous production level **do**
        i. identify the direct buyers of the affected node
        ii. calculate the fraction of inputs the buyer sources from the affected node
        iii. reduce the output level of the buyers by the fraction calculated in ii.
        iv. update the set of nodes with lower production level than in previous iteration
    b. **end for**
3. **end while**
4. **return** output level of each node, $h_j(T)$, where $T$ is the last iteration.

**Appendix C: Details on the network and the network construction**

We briefly describe how the 22,938 nodes are distributed over the considered business premises type categories. Most nodes belong to the primary production sector (15953) consisting of mostly different types of pig farms, but also wholesale traders and animal transporters. The second largest category is constituted by the different supermarket stores (3525), followed by the meat processors (2687) and slaughterhouses (749). The distribution centers of the supermarket chains are the smallest node group (24) illustrating their role as potential bottlenecks.

We quantify the transfers between farms and to slaughterhouses in yearly numbers of pigs, whereas the other transfers are counted in kg of pork meat per year. It is considered that on average 70kg of meat are recovered from an individual pig that is slaughtered.

For the network construction we used the following 5 data sources.
- **Animal movement data.** Records of transfers of living pigs within, out of and into Austria. The data is available daily over 2 years (2019-2020). All movements between farms and slaughterhouses are listed, as well as transfers to and from other countries. The data set is provided by the Austrian health ministry in strictly anonymized form. For the empirical analysis we use the available data from July to December 2020 representing the time span from birth to death of a pig.

- **Supermarket chain delivery data.** Daily incoming deliveries of pork meat products from suppliers (slaughterhouses, meat processors) to distribution centers and daily deliveries of pork meat products from distribution centers to supermarkets. The data includes the volumes (in kg) for the respective product types delivered and spans several months in 2021 for a large supermarket chain with a market share at the magnitude of one quarter. For other supermarket chains this data is imputed with publicly available data (locations of supermarket stores and distribution centers) and the imputation method described below.
- **Population census data.** Population data for each district is obtained from the Austrian statistical office.
- **List of meat processors and slaughterhouses.** All meat processors and slaughterhouses licensed in Austria and their geographical locations; obtained from the Austrian statistical office.
- **Production volumes of slaughterhouses and meat processors.** The volumes are estimated for the largest slaughterhouses and meat processors covering around 90% of the production. The estimates were provided by the agricultural ministry for 2019.
- **Distribution center and supermarket locations.** The geographical positions of distribution centers and supermarkets are publicly available for all major supermarket chains in Austria.

To impute missing supply links in the network we use the following imputation rules:
- Supply relations between distribution centers and supermarkets that were not available are inferred from attributing the supermarket stores to the distribution centers of the respective supermarket chains according to federal state membership. The weights for the transactions are stochastically imputed by drawing link weights from the known empirical distribution of distribution centers to supermarket delivery volumes.
- The production volumes of small meat processors for which no estimates of the production volumes are available are assumed to follow an exponential distribution that is parameterized such that the remaining 10% of the production volume in Austria is attributed.
- We impute the links and the supply volumes between slaughterhouses and meat processors and from meat processors to distribution centers where not available with a geographical distance-based heuristic, i.e., meat processors buy slaughtered pigs from geographically close slaughterhouses and large slaughterhouses can supply further away meat processors. We assume the same heuristic for the links from meat processors to distribution centers where they are not known.